\documentclass{jetpl}
\twocolumn

\usepackage{ dsfont }
\usepackage[]{breqn}
\usepackage{tikz}
\usepackage{graphicx}
\usepackage{pinlabel}

\def\uqslN{$\mathcal{U}_q(sl_N)$}

\def\uqsltwo{$\mathcal{U}_q(sl_2)$}

\def\Tr{{\rm Tr}\,}

\def\R{$\mathcal{R}$}
\def\r{\mathcal{R}}
\def\P{\mathcal{P}}
\def\M{\mathcal{M}}
\def\K{\mathcal{K}}
\def\L{\mathcal{L}}
\def\A{\mathcal{A}}
\def\W{\mathcal{W}}
\def\l{\lambda}
\def\nn{\nonumber}

\def\bc{\begin{center}}
\def\ec{\end{center}}

\newcommand{\prin}[2]{[#2]_{#1}}
\newcommand{\hev}[2]{\theta_{#1}\!\left(#2\right)}

\newcommand{\pic}[2]{\raisebox{0.0\height}{\includegraphics[scale=#2]{#1}}}

\newcommand{\picb}[2]{\raisebox{-0.45\height}{\includegraphics[scale=#2]{#1}}}
\newcommand{\picc}[2]{\raisebox{0.35\height}{\includegraphics[scale=#2]{#1}}}

\def\hopf{\picc{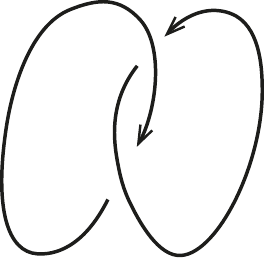}{.700}}
\def\trefoil{\pic{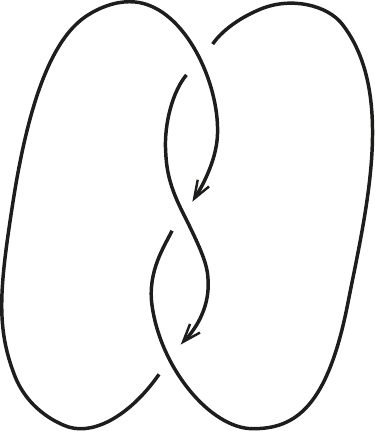}{.700}}
\def\hopfcut{\picc{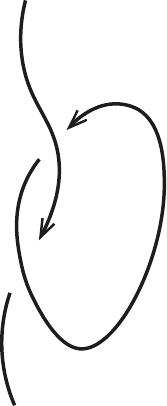}{.700}}
\def\hopfcuto{\pic{Pictures/hopflinkcut}{.700}}
\def\trefoilcut{\pic{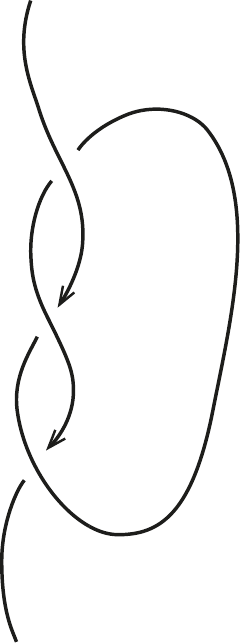}{.700}}

\def\Rone{\pic{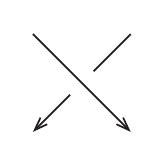}{.600}}
\def\Rtwo{\pic{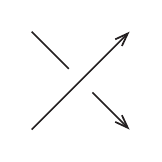}{.600}}
\def\Rthree{\pic{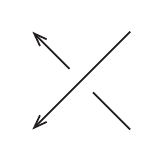}{.600}}
\def\Rfour{\pic{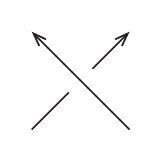}{.600}}
\def\Rfive{\pic{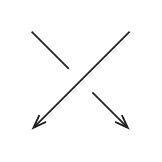}{.600}}
\def\Rsix{\pic{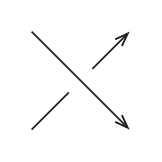}{.600}}
\def\Rseven{\pic{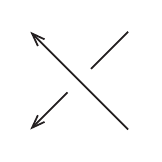}{.600}}
\def\Reight{\pic{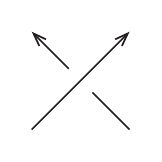}{.600}}

\def\Mone{\pic{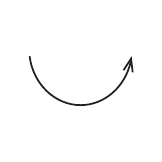}{.600}}
\def\Mtwo{\pic{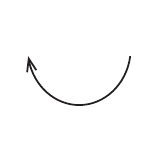}{.600}}
\def\Mthree{\pic{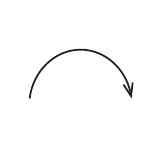}{.600}}
\def\Mfour{\pic{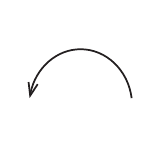}{.600}}

\def\loopone{\picb{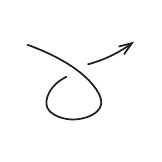}{.600}}
\def\loopthree{\picb{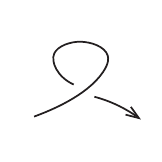}{.600}}

\def\Reione{\picb{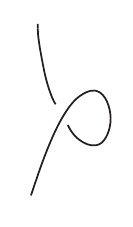}{.600}}
\def\Reitwo{\picb{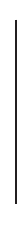}{.600}}
\def\Reithree{\picb{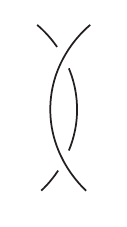}{.600}}
\def\Reifour{\picb{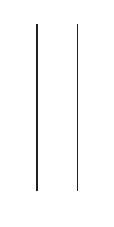}{.600}}
\def\Reifive{\picb{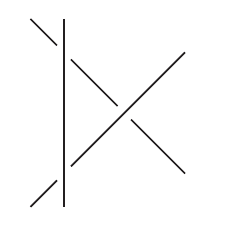}{.600}}
\def\Reisix{\picb{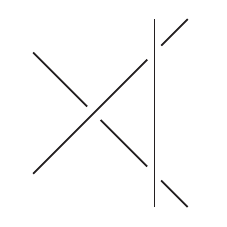}{.600}}

\def\Monea{\picb{Pictures/M1}{.600}}
\def\Mtwoa{\picb{Pictures/M2}{.600}}
\def\Mthreea{\picb{Pictures/M3}{.600}}
\def\Mfoura{\picb{Pictures/M4}{.600}}
\def\Ronea{\picb{Pictures/R1}{.600}}
\def\Rfoura{\picb{Pictures/R4}{.600}}
\def\Reighta{\picb{Pictures/R8}{.600}}
\def\kkk{\picb{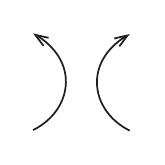}{.600}}

\def\auplong{\pic{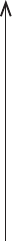}{.600}}
\def\adown{\pic{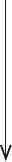}{.600}}
\def\adotdown{\pic{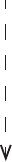}{.600}}
\def\aleft{\pic{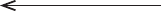}{.600}}
\def\aright{\pic{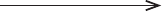}{.600}}

\lat

\title{Overview of knot invariants at roots of unity}

\rtitle{Overview of knot invariants at roots of unity}

\sodtitle{Overview of knot invariants at roots of unity}

\author{{ Liudmila Bishler$^{a,b,c}$}\vspace{1cm}}

\rauthor{{ L. Bishler}}

\sodauthor{{ L. Bishler}}

\address{$^a$ {\small {\it Lebedev Physics Institute, Moscow 119991, Russia}}\\
$^b$ {\small {\it Institute for Information Transmission Problems, Moscow 127994, Russia}}\\ $^c$ {\small {\it bishlerlv@lebedev.ru}}}

\abstract{We discuss different invariants of knots and links that depend on a primitive root of unity. We clarify the definitions of existing invariants with the Reshetikhin-Turaev method, present the generalization of ADO invariants to \uqslN{} and highlight the connections between different invariants.}

\PACS{ }

\begin{document}

\maketitle

\section{Introduction}

Celebrated Jones polynomial $J^{\K}(q)$ discovered in 1984 by V.~Jones \cite{Jones} is a one-variable polynomial invariant of knots and links. It was originally defined with  skein-relations, which offer a constructive method of calculation of Jones polynomials. Skein relations (\ref{skein}) connect Jones polynomials of three knots that differ in one crossing (\ref{skein0}). 
\begin{equation}
    J^{\mathcal{K}}(q)) - q^{-2}J^{\mathcal{K}'}(q) = (q-q^{-1})J^{\mathcal{K''}}(q).
    \label{skein}
\end{equation}

\begin{equation}
    \K\,\, = \Reighta{},\quad \K^'\, = \Rfoura{}, \quad \K^{''}= \kkk{}
    \label{skein0}
\end{equation}

Shortly after the definition of Jones polynomial two important  discoveries were made by E.~Witten and N.~Reshetikhin and V.~Turaev. E.~Witten \cite{Witt} found a quantum field theory --- Chern-Simons theory with gauge group $SU_2$ --- that allowed to construct observables (Wilson loop averages) that coincide with Jones polynomials, i.e. he offered a physical definition of a mathematical object. N.~Reshetikhin \cite{Resh} and V.~Turaev \cite{Tur}  on the other hand discovered a new method of calculation of knot invariants with special operators --- \R-matrices. They connected Jones polynomials with universal \R-matrix in fundamental representation of quantized universal enveloping algebra \uqsltwo{}. The Reshetikhin-Turaev (RT) method also allowed to define colored Jones polynomials calculated with \R-matrices in other representations of \uqsltwo{}. 

These results were later generalized to HOMFLY-PT polynomials \cite{HOMFLY, PT}, Chern-Simons theory with gauge group $SU_N$ and universal \R-matrix in representations of \uqslN{}.  



 \uqslN{} is an associative algebra with generators $E_i$, $F_i$, $K_i=q^{h_i}$ ($i=1,\dots,N-1$), that satisfy the relations 
  \begin{equation}
\begin{array}{ll}
\ K_i E_j =q^{a_{ij}} E_j K_i, & [K_i, K_j] = 0,
\\ \
K_i F_j = q^{-a_{ij}} F_j K_i, & [E_i, F_j] = \delta_{ij} \frac{K_i-K^{-1}_i}{q-q^{-1}}.
\end{array}
\label{defsln}
\end{equation}
The universal \R-matrix is the following
\begin{equation}
\mathcal{R}_u = P q^{\sum_{i,j} a^{-1}_{i,j} h_i \otimes h_j} \overrightarrow{\prod_{\beta \in \Phi^{+}}} {\rm exp}_{q}     \left ( (q - q^{-1}) E_{\beta} \otimes F_{\beta}   \right),
\label{Rsln}
\end{equation}
where $P (x \otimes y) = y \otimes x$, $\Phi^{+}$ -- positive roots, ${\rm exp}_q A = \sum_{m = 0}^{\infty} \frac{A^m}{[m]_q !} q^{m(m-1)/2}$, $[m]_q = (q^m-q^{-m})/(q-q^{-1})$.

\bigskip

 Chern-Simons (CS) theory is the three-dimensional quantum field theory with the action
 \begin{equation}
  \mathcal{S}_{CS} =   \frac{k}{4\pi} \int_{\mathcal{M}} \Tr \left(\mathcal{A}\wedge d\mathcal{A}+\frac{2}{3}\mathcal{A}\wedge\mathcal{A}\wedge\mathcal{A}\right).
\label{cs}
 \end{equation}
 Non-zero correlators in CS theory are the correlators of a special type --- Wilson loop averages. When the gauge group of the theory is $SU_N$, they coincide with HOMFLY-PT polynomials $H_R^{\cal K}(q,A)$.
\begin{equation}
    H_R^{\cal K}(q,A) =
\frac{1}{d_R(N)}\left< {\rm Tr}_R P\exp\left(\oint_{\cal K} {\cal A}\right)\right>
_{{\rm CS}(N,k)}.
\end{equation}
They depend on a contour ${\cal K}$ (a knot or a link), the rank $N-1$ of the gauge group $SU_N$,
its representation (corresponding to a Young diagram) $R$, on quantum dimension $d_R(N)$
and the Chern-Simons coupling constant $k$. This average is a polynomial in variables $q = \exp\left(\frac{2\pi i}{N+k}\right)$ and $A=q^N$. It was shown \cite{Marino} that CS theory is gauge invariant when the coupling constant $k$ is an integer, which means that $q$  is a root of unity. That is why invariants at roots of unity attract additional attention. 

The obvious approach to get invariants at roots of unity is to substitute variables in HOMFLY-PT $H_R^{\K}(q,A)$ \cite{KonMor} and Jones $J_{[r]}^{\K}(q)$ polynomials. 

There also exist the invariant $\langle K\rangle_{m,N}$ defined by R.~Kashaev \cite{Kasnew,Kas,Kas96} with \R-matrix that depends on a variable $\omega$ (that is a primitive $N$-th root of unity) and an integer parameter $m$. The Kashaev invariants are not connected with quantum algebras, however they coincide with colored Jones polynomials. 

Another possibility to construct invariants at roots of unity emerges when we consider representations of \uqslN{} when the parameter of quantization is a root of unity. In this case new types of representations with parameters $\l$ emerge, which allow one to construct \R-matrices with parameters and to define new invariants of knots and links at roots of unity. The resulting invariants are ADO \cite{ADO} or colored Alexander invariants \cite{Murakami} $\Phi_m^{\L}(q, \lambda)$ for \uqsltwo{}. In this case $q$ is $2m$-th root of unity and $\lambda$ is an arbitrary parameter. ADO invariants coincide with Alexander polynomials $\mathcal{A}(q) = H_{[1]}^{\K}(q,A=1)$ when $q$ is $4$-th root of unity. They are also connected with Jones polynomials. 

The new result that we want to highlight in this letter is the generalization of ADO invariants to \uqslN{}. These are the invariants $\mathcal{P}^{\L}_{m,N}(q, \l_i)$ (\ref{Pinv}) \cite{mila1} of knots and links, which are associated with nilpotent representations with parameters of \uqslN{} at roots of unity ($q^{2m}=1$). They depend on a set of parameters $\{\l_1, \dots, \l_{N-1}\}$ and are connected with Alexander and HOMFLY-PT polynomials. 
 ADO invariants and invariants $\mathcal{P}^{\L}_{m,N}(q,\l_i)$ are defined with the modified version of the Reshetikhin-Turaev method, which requires the introduction of a special normalization coefficient (\ref{norm}) that we present in this letter. 

The schematic correspondence between invariants described above is on Fig.1.

\begin{table}[h!]
    \centering
    \begin{tabular}{ccccc}
  $ \underset{{q^N=1}}{\langle K \rangle_{m,N}}$    &  $\overset{m=r}{\aright{}}$ & $J_{[r]}^{\K}(q)$ &  $\overset{N=2}{\aleft{}}$ & $H_R^{\K}(q,A=q^N)$ \\
 &  &{ \labellist \pinlabel \rotatebox{90}{\scriptsize{$\l=q^{m-1}$}} at -10 30 \endlabellist \auplong{} }& &  { \labellist \pinlabel \rotatebox{90}{\scriptsize{$\l_i=q^{m-1}$}} at -10 30 \endlabellist \auplong{} }  \\
 & &$\underset{q^{2m}=1}{{\Phi_{m}^{\mathcal{L}}\left(q,\lambda \right)}}$ & $\overset{N=2}{\aleft{}}$ & $\underset{q^{2m}=1}{\mathcal{P}^{\L}_{m,N}\left(q,\l_i\right)}$ \\
  &  & { \labellist \pinlabel \rotatebox{90}{\scriptsize{$q^4=1$}} at -10 23  \endlabellist \adown{} } & &  {\labellist \pinlabel{\scriptsize{\cite{mila1}}} at -13 25 \endlabellist \adotdown{} } \\
  & & $\mathcal{A}\left(\l \right)$& & $\prod_i \mathcal{A}(\l_i)$ \\
    \end{tabular}
    \caption{1. Correspondence between the invariants at roots of unity}
    \label{table}
\end{table}

The aim of this letter is to clarify the definition of different invariants at roots of unity and establish relations between them. The structure of the letter is the following. We start with the Reshetikhin-Turaev (RT) method (sec. 2) that is used to define all invariants considered in this letter. Then we discuss the representation structure of \uqsltwo{} for different values of $q$ (sec. 3). We define ADO invariants (sec. 4) and their generalization (sec. 5) and discuss the modifications of RT method that are necessary in order to define them. Finally we consider the notion of long knots and define Kashaev invariant (sec. 6). The new results that we present in this letter are in sec. 5.

\section{Reshetikhin-Turaev method}

The Reshetikhin-Turaev (RT) method \cite{Kas,RT1, MorSmir} allows one to define colored invariants of knots and links \cite{mut1, mut2}. There are also modifications of this method that allow to conduct the calculations more efficiently in some cases \cite{RTMM1,RTMM2}. In this section we discuss the most general version of the RT method and follow the description from \cite{MorSmir}.

The RT method is based on the use of a two-dimensional oriented projection of a knot or a link on a plane with a fixed direction, which is a diagram of a knot or a link. The diagram shows which thread is above the other in each crossing. Then a diagram is broken down into the elements that play the role in the construction of knot invariant: crossings and turning points (relative to the selected direction). There are eight types of crossings and four types of turning points (Fig.2).

\begin{figure}[h!]
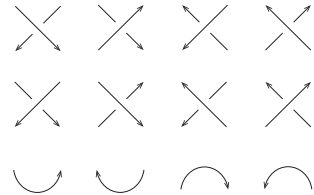

\bc
{\Rone{}  \Rtwo{}  \Rthree{} \Rfour{}}  \\
{ \Rfive{}  \Rsix{} \Rseven{}  \Reight{}}\\
{\Mone{}  \Mtwo{}  \Mthree{}  \Mfour{}}
\ec
\caption{2. Different types of crossings and turning points}
\end{figure}

 \noindent All turning points and crossings can be expressed just with the operators $\r$, $\M$ and $\overline{\M}$:
\begin{align}
{\Ronea} = \,\, \r  \quad {\Monea} = \,\, \M \quad {\Mfoura} = \,\, \overline{\M}
\end{align}

\begin{figure}[h!]
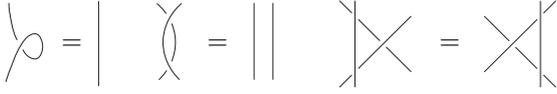

    \bc
    {\Reione} $=$ {\Reitwo} \quad  {\Reithree} $=$ {\Reifour} \quad  {\Reifive} $=$ {\Reisix}
    \ec
    \caption{3. Reidemeister moves}
    \label{rei}
\end{figure}

\noindent Operators $\r$, $\M$ and $\overline{\M}$ satisfy the equations that come from Reidemeister moves (Fig.3):
\begin{align}
   & {\rm Tr}_2 (I \otimes \W) \, \r  = q^{\Omega} I  \label{1st}\\
   &  \r_1 \r_2 \r_1 = \r_2 \r_1 \r_2 \label{YB},
\end{align}
where $\r_1 = \r\otimes I$, $\r_2 = I\otimes \r$, $\W = \M \overline{\M}$, $I$ --- identity operator. Equations (\ref{1st}) and (\ref{YB}) fix only \R{} and $\W$ operators, so there is some freedom in definition of $\M$ and $\overline{\M}$. 
Different choices of operators \R{} and $\W$ can produce different types of invariants. Colored Jones and HOMFLY-PT polynomials are associated with universal \R-matrix (\ref{Rsln}) in representations of \uqsltwo{} and \uqslN{} correspondingly, ADO invariants and their generalization are associated with universal \R-matrix in nilpotent representation with parameters of \uqsltwo{} and \uqslN{} at roots of unity. The Kashaev invariant is based on a different \R-matrix (\ref{rKashaev}). 

Coefficient $q^{\Omega}$ in eq.(\ref{1st}) is called the framing coefficient. It emerges when we consider a knot made out of a ribbon. In this case the first Reidemeister move resolves with a coefficient. In topological framing, which we use in definition of ADO invariants, matrices \R{} and $\W$ satisfy the equation 
\begin{equation}
    {\rm Tr}_2 (I \otimes \W) \, \r  = I.
\end{equation}
The polynomial invariant of a knot or a link is defined as a contraction of all operators associated with elements of a particular diagram.

If we choose universal \R-matrix calculated for representation $R$ of \uqslN{}, this method gives us unreduced HOMFLY-PT polynomial $\mathcal{H}_{R}^{\mathcal{K}}$. We can also define reduced polynomials  $H^{\mathcal{K}}_{R} = \mathcal{H}_{R}^{\mathcal{K}}/\mathcal{H}_{R}^{\circ}$, where $\mathcal{H}_{R}^{\circ}$ is the unreduced polynomial of unknot.

\begin{figure}[h!]
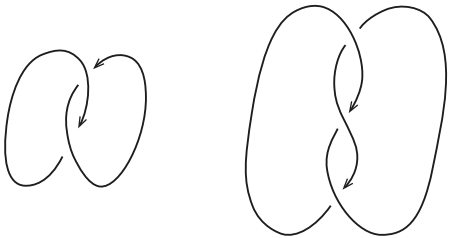

\bc
{\hopf{}} {\quad \quad \quad } {\trefoil{}}
\caption{4. Diagrams of hopf link and trefoil knot in the braid form}
\ec
\end{figure}

It is convenient to use diagrams of knots and links in the form of braids (Fig.4), that exist for any knot or link. In this case the definition of invariants can be reformulated in terms of Markov trace $\Tr_q$ (quantum trace) and operator $\W=\M \overline{\M}$, which is known as the weight matrix. 

\begin{equation}
    \mathcal{H}_{R}^{\mathcal{K}} = \Tr_q \, \prod_{i} \r_i = \Tr\, \overbrace{ \W\otimes\dots \otimes \W}^{s}  \prod_{i} \r_i, 
    \label{homfly}
\end{equation}
where $s$ is a number of strands in a braid, the product includes all crossings in the braid.

\section{Representations of $\mathcal{U}_q(sl_2)$ at roots of unity}
\uqsltwo{} is generated by elements $e$, $f$, $k = q^h$ and $k^{-1} = q^{-h}$ that satisfy the relations 

\begin{align}
    k k^{-1} & = k^{-1} k = 1,\,\,\,\,\,
    k e k^{-1} = q^2 e,  \label{algebra}\\
    k f k^{-1} & = q^{-1} f, \,\,\,\,\,
    [e,f] = ef-fe  = \frac{k-k^{-1}}{q-q^{-1}}. \nonumber
\end{align}

\noindent The universal \R-matrix is the following
\begin{equation}
    \r = P\,  q^{h\otimes h/2} \sum_{m=0}^{\infty} \frac{q^{m(m+1)/2} (1-q^{-2})^m}{[m]_q!}\, e^m \otimes f^m
\end{equation}
and the corresponding weight matrix coincides with operator $k$: $\W=k$.

When $q$ is not a root of unity the irreducible finite dimensional representations $L_r$ of the algebra \uqsltwo{} are symmetric representations enumerated with Young diagrams that consist of one row $[r]$. $L_r$ are the representations with the highest and the lowest weights, which act on a vector space $\mathcal{V}_{r+1}$ of dimension $r+1$ with basis vectors $v_i$, $i = \{0, \dots, r \}$, where $v_0$ and $v_r$ are the highest and lowest weight vectors correspondingly. 
\begin{align}
    L_r(k) v_i &= q^{r-2i}v_i, \nn & \\
    L_r(e) v_i &= \,[i]_q[r-i+1]_q v_{i-1},\label{repnot} &  L_r(e) v_0 = 0, \\
    L_r(f)v_i &= \, v_{i+1}, \nn &  L_r(f)v_{r} = 0,
\end{align}
where $[x]_q = (q^x-q^{-x})/(q-q^{-1})$ is a quantum number, $\delta_{ij}$ is Kronecker delta.
In this case the highest weight $L_r(k) v_0 = \l\, v_0$ is fixed $\l = q^r$. The condition that fixes the weight emerges when one builds a Verma module starting with an eigenvector $v_0$ of operator $k$ that satisfies $e\, v_0 = 0 $. One gets the other vectors of the Verma module acting on $v_0$ with the operator $f$: $f \, v_0 = v_1$, $f^2 \, v_0 = v_2$, $\dots$, $f^n \, v_0 = v_n$. Then we look for an invariant subspace with the condition $e\,v_{r+1}=0 $, which is the following:
\begin{equation}
    [r+1]_q(\lambda q^{-r}-\lambda^{-1}q^{r})=0.
    \label{weightcond}
\end{equation}
This condition fixes the weight $\l$ only if $[r+1]\neq 0$, which means that when $q$ is a root of unity, there exist representations, where the weight is arbitrary. 
\bigskip

Let $q$ be a primitive root of unity of degree $2m$, which means that there is no $p<2m$ so that $q^p=1$. In this case operators $e^m$, $f^m$ and $k^m$ are central, which comes directly from the defining relations (\ref{algebra}). The centrality of $k^m$ results in the fact that the weight of representations of dimension $m$ is not fixed and is a parameter of representations. The fact that $e^m$ and $f^m$ are central is the reason why new types of representations emerge in this case: cyclic and semi-cyclic. 

There are four types of irreducible  representations (any irreducible representation of \uqsltwo{} is finite-dimensional at roots of unity): 
\begin{enumerate}
    \item representations $L_{r}$ (\ref{repnot}) for $r \leq m-2$,
    \item cyclic $U_{m}^{a,b,\l}$,
    \item semi-cyclic $V_{m}^{a,\l} = U_{m}^{a,0,\l}$ or $V_{m}^{b,\l}=U_{m}^{0,b,\l}$, 
    \item nilpotent representations $W_{m}^{\l}=U_{m}^{0,0,\l}$. 
\end{enumerate}

The last three representations have the same dimension $m$ and can be described with the following operators acting on a $m$-dimensional vector space $\mathcal{V}_m$  with basis $v_i$, $i = 0,1, \dots, m-1$.
\begin{align}
     U^{\lambda, a, b}_{m}(k)& v_i = q^{-2 i} \lambda v_i  \\
    U^{\lambda , a, b}_{m} (e) & v_i= \left( ab + [i]_q \frac{\lambda q^{1-i}-\lambda^{-1}q^{i-1}}{q-q^{-1}}\right)v_{i-1}, \,\,\, i>0 \nn \\
      U^{\lambda , a, b}_{m}(e)& v_0 = a v_{m-1}, \nn \\
      U_{m}^{\lambda, a, b}(f) & v_i = v_{i+1}, \,\,\, i<m-1 \nn\\
      U^{\lambda , a, b}_{m} (f)& v_{m-1}= b v_0, \nn 
\end{align}
where $a$, $b$, $\lambda$ are arbitrary complex numbers such that $\lambda \neq 0$. One can check that these operators satisfy the defining relations (\ref{algebra}) of $\mathcal{U}_q(sl_2)$.

Representations $W_m^{\l}$ produce non-trivial \R-matrices and allow one to define invariants of knots and links that are known as ADO or colored Alexander invariants.

\section{ADO or colored Alexander invariants}
ADO invariants of links were defined by Akutsu, Deguchi and Ohtsuki in \cite{ADO}. ADO invariants of knots and links can be defined with Reshetikhin-Turaev method, which is applied to $(1,1)$-tangles --- knot and links with one cut line (Fig.5). The consideration of tangles instead of knots and links is possible because of the existing one-to-one correspondence between them \cite{tangles}. It is an important step that allows one to calculate non-zero invariants. Invariants, that are calculated with knots and links directly are all equal to zero, because of the properties of Markov trace in representations $W_{m}^{\l}$.

\begin{figure}
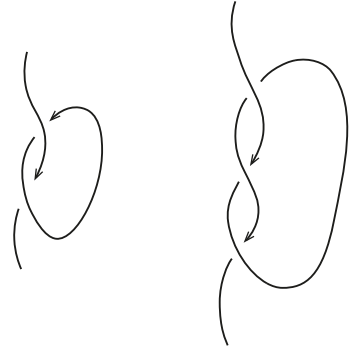

\bc
   \hopfcut{} \quad \quad\quad\quad \trefoilcut{}
   \ec
   \caption{5. $(1,1)$-tangles of hopf link and trefoil knot}
\end{figure}

Let us now define ADO invariants. Two important modifications of the RT method have to be made. First of all, we have to redefine Markov trace in the following way:
\begin{equation}
  \boxed{  {\rm Tr^*}_q \dots = {\rm Tr}\,\, I \otimes \overbrace{ \W \otimes \dots \otimes \W}^{s-1} \dots, }
    \label{trdef}
\end{equation}
i.e. omit one weight matrix, that is associated with a cut line. Then the normalization coefficient of polynomials (unreduced polynomial of unknot) equals a classical dimension of a representation.

This procedure makes one choose the line to cut and brings asymmetry into the definition of the invariant. That is why one also needs to introduce a normalization coefficient.  The coefficient that was calculated in \cite{ADO} up to normalization coefficient $q^{m}$ is the following
\begin{equation}
\boxed{
    \Xi_{m}^{sl_2}(\l_1) =\prod_{i=0}^{m-2} \{ \l_1 q^{-i} \},}
    \label{ADOnorm}
\end{equation}
where $\{x\} = x-x^{-1}$, $\l_1$ is a color of an open component. 

Now we can define ADO invariant $\Phi_{m}^{\L}(\l_1, \dots)$:

\begin{equation}
    \boxed{
    \Phi_{m}^{\L}(\l_1, \dots) = \frac{ {\rm Tr}\,\, I \otimes \overbrace{ \W \otimes \dots \otimes \W}^{s-2}  \prod_{i} \r_i}{ \Xi_{m}^{sl_2}(\l_1)}.
    }
    \label{ADOdef}
\end{equation}

In this definition \R{} is the universal \R-matrix, calculated for representation $W_m^{\l}$ of \uqsltwo{} at roots of unity. In general it depends on two colors $\l_1 = q^{\mu_1}$ and $\l_2=q^{\mu_2}$:

\begin{dmath}
   \r_{m}(v_i^{\mu_1} \otimes v_j^{\mu_2}) =   \sum_{n=0}^{m-1} q^{\mu_2(m-1-i+n)-\mu_1(j+n)+2(i-n)(j+n)+\frac{n(n-1)}{2}} \\ \frac{(q-q^{-1})^n}{[n]!}   {[i-n+1;n]} {[\mu_1-i+n;n]} \, (v_{j+n}^{\mu_2}\otimes v_{i-n}^{\mu_1}).
   \label{R2mix}
\end{dmath}
and 
\begin{equation}
    \mathcal{W}_{m} \, v_i^{\mu} =q^{-\mu (m-1)-2i} v_i =  q^{-\mu\, m} k v_i^{\mu}.
    \label{W2}
\end{equation}

\bigskip
ADO invariants $\Phi_{m}^{\L}$ are connected with Alexander and Jones polynomials. For simplicity let us define ADO polynomials of knots and one-colored links:
\begin{equation}
    \hat{\Phi}_{m}^{\L}(\l) = \Phi_{m}^{\L}(\l) \,  \Xi_{m}^{sl_2}(\l)
    \label{ADOA}
\end{equation}
then 
\begin{equation}
    \hat{\Phi}_{2}^{\K}(\l) = \mathcal{A}^{\K}(q=\l), 
    \label{ADOA1}
\end{equation}
i.e. ADO polynomials for 4-th root of unity coincide with Alexander polynomials of knots and ADO invariants for for 4-th root of unity coincide with multivariable  Alexander polynomials of links
\begin{equation}
    \Phi_2^{\L}(\l_1, \dots) = \Delta^{\L}(\l_1,\dots),
\end{equation}
and that is why ADO invariants are also called colored Alexander invariants.

The connection with Jones polynomials is the following:
\begin{equation}
      \hat{\Phi}_{m}^{\L}(\l = q^{m-1}) =\left. J_{[m-1]}^{\L}(q)\right|_{q^{2m}=1},
      \label{ADOJ}
\end{equation}
where $ J_{[m-1]}^{\L}(q)$ are reduced Jones polynomials in representation $L_{m-1}$. It follows from the fact that representations $W_{m}^{\l}$ coincide with representations $L_{m-1}$ when we choose the correct value of the weight $\l = q^{m-1}$.

Recent study by S.Willetts \cite{Sonny} showed that ADO invariants and colored Jones polynomials can be generalized with the unified knot invariant that contains both invariants: ADO and Jones. And there exists a map that allows to get ADO invariants from colored Jones polynomials. 

\section{Generalization of ADO invariants to \uqslN{}}

Jones polynomials were generalized to HOMFLY-PT, and similarly ADO invariants can be generalized to invariants $\P^{\L}_{m,N}(\l_i^{(j)})$ associated with  representations of \uqslN{} (\ref{defsln}) at roots of unity.

Let $q$ be a primitive root of unity of degree $2m$. In this case the representation structure of \uqslN{} \cite{Arnaudon} is very similar to representation structure of \uqsltwo{}, that we discussed before. Operators $K^m_i$ are central and there exist representations of dimension $m^{N(N-1)/2}$ with arbitrary weights --- nilpotent representations $W_{m,N}^{\l_i}$ with $N-1$ parameters $\l_i$. There are also cyclic and semi-cyclic representations because $E_i^m$ and $F_i^m$ are central, but these representations do not produce non-trivial invariants of knots and links \cite{mila1}. Representations $W_{m,N}^{\l_i}$ are associated with non-trivial invariants that we denote $\mathcal{P}^{\L}_{m,N}(\l_i^{(j)})$.

The definition of invariants $\mathcal{P}^{\L}_{m,N}(\l_i^{(j)})$ repeats the definition of ADO invariants. We color components of a link with $l$ representations $W_{m,N}^{\l_i^{(j)}}$ ($j=1,\dots, l$) with sets of parameters $\l_i$ ($i=1,\dots, N-1$), make a projection of the link and cut one line of the projection.  We then apply RT method to $(1,1)$-tangles and get polynomials $P_{m,N}^{\L}(\l_{i}^{(j)})$. As operators \R{} and $\W$ in RT method we use  the universal \R-matrix (\ref{Rsln}) and operator $\mathcal{W}$, calculated for representations $W_{m,N}^{\l_i}$. Operator $\W$ is the following
\begin{equation}
    \mathcal{W} = q^{2 h_{\rho}}, \quad h_{\rho} = \frac{1}{2} \sum_{\alpha \in \Phi^{+}} h_{\alpha}.
\end{equation}

\begin{figure}[h!]
\bc
{\labellist
 \pinlabel{$\l^{(1)}$} at -7 110 
 \pinlabel{$\l^{(2)}$} at 60 50 
\endlabellist
\hopfcuto}
\ec
\caption{6. Colored $(1,1)$-tangle corresponding to hopf link}
\label{coloredhopf}
\end{figure}

 We also need to normalize the polynomials $P_{m,N}^{\L}(\l_{i}^{(j)})$ in order to restore the symmetry between all threads in a link. If the color of an open component is $\l^{(1)}$ (Fig.6) the normalization coefficient $\Xi_{m,N}(\l^{(1)})$ is the following

\begin{equation}
\boxed{
   \Xi_{m,N}(\l^{(1)}) =  \prod_{\alpha \in \Phi^{+}_N} \xi_m( \l^{(1)}_\alpha q^{|\alpha|}),
   } \label{norm}
\end{equation}
where $\alpha$ are positive roots $\Phi^{+}_N$ of $sl_N$,  $\alpha=\sum_{k=i}^j\alpha_k$ ($i\leq j < N $), where $\alpha_k$ are simple roots of $sl_N$, $|\alpha|=j-i$. Definition of $\xi_m(\lambda)$ repeats the definition of normalization coefficient of ADO invariants (\ref{ADOnorm}):
\begin{equation}
    \xi_m(\lambda)=\prod_{i=0}^{m-2} \{\lambda q^{-i}\}.
\end{equation}

Now we can define invariants $\P_{m,N}^{\L}(\l^{(j)}_i)$ at roots of unity with parameters $\l_i^{(j)}$ ($i=1,\dots,N-1$, $j=1,\dots,l$, $l$ -- number of components in a link), $\l^{(1)}_i$ is the color of an open component:
\begin{equation}
\boxed{
    \P_{m,N}^{\L}(\l^{(j)}_i) = {P_{m,N}^{\L}(\l_{i}^{(j)}) \over \Xi_{m,N}(\l^{(1)}_i)}.}
    \label{Pinv}
\end{equation}

\bigskip

Invariants $\P_{m,N}^{\L}(\l^{(j)}_i)$ coincide with HOMFLY-PT polynomials in representations $R_{m,N}$ corresponding to the Young diagrams $[(N-1)(m-1), (N-2)(m-1), \dots, (m-1)]$ when the parameters $\l_i^{(j)}$ coincide with the highest weights of representation $R_{m,N}$:

\begin{equation}
 \left.   P_{m,N}^{\mathcal{L}}\,(\lambda_i^{(j)} = q^{m-1})\right|_{q^{2m}=1} = \left. H^{\L}_{R_{m,N}} (A=q^N,q)\right|_{q^{2m}=1}.
 \label{PH}
 \end{equation}
 
\noindent They are also connected with Alexander polynomials, however these connections are not as simple as in case of ADO invariants. They are listed in \cite{mila1}. For example  for $N=3$:
\begin{equation}
     P^{\K}_{2,3}\,(\lambda_1, \l_2=1) = \A^{\K} (\lambda^{2}_1),
     \label{PA1}
\end{equation}
\begin{equation}
     P^{\K}_{3,3}\,(q,\lambda_1, \l_2=1) =\A^{\K} (\lambda_1) \A^{\K} (\lambda^{3}_1).
     \label{PA2}
\end{equation}
Invariants $\P_{m,N}^{\L}(\l^{(j)})$ coincide with ADO invariants $\Phi_{m}^{\L}(\l^{(1)}, \dots)$ when $N=2$.

\section{Kashaev invariant}
There exists another type of knot invariant, defined specifically for a root-of-unity variable, which is not based on representations of quantum algebras. In this section we discuss the Kashaev knot invariant $\langle K \rangle_{m,N} $. 

The Kashaev invariant of knots was defined by Rinat Kashaev in his work \cite{Kas96} for long knots. It is based on \R-matrix, obtained from solutions of pentagon identity, which depends on a root-of-unity variable $\omega$ and two integer spectral parameters ($m$ and $n$) that are associated with two colors of two strands in a crossing.  For the definition of the invariant Kashaev uses the RT method applied to $(1,1)$-tangles --- two-dimensional projections of long knots. Long knots are 3-dimensional analogs of $(1,1)$-tangles. By definition long knot is an embedding $f$: $\mathds{R} \longrightarrow \mathds{R}^3$ and there exist $a$, $b$ $\in$ $\mathds{R}$ such that $f(t) = (0,0,t)$ for any $t<a$ or $t>b$. Calculating invariants of long knots allows to avoid the problem with normalization coefficient.

The steps to define the Kashaev invariant are the following. First of all, we fix a primitive root of unity $\omega$ of order $N$, color the threads with integer numbers $n_i$: $0 \leq n_i < N$, make a two-dimensional projection, place \R-matrices and turning point operators according to the rules below and sum over all indices. The invariance of the resulting sum was shown in \cite{Kas}.

\begin{align}
     {\labellist   
   \pinlabel{\tiny{$m$}} at 8 18
   \pinlabel{\tiny{$n$}} at 35 18
   \pinlabel{\tiny{$k$}} at 6 0
   \pinlabel{\tiny{$l$}} at 36 0
   \pinlabel{\tiny{$j$}} at 6 45
   \pinlabel{\tiny{$i$}} at 36 45
   \endlabellist \Reighta{}}  & = \,  \langle i,j|r(m,n)|k,l\rangle  \\
      {\labellist
        \pinlabel{\tiny{$i$}} at 6 17
    \pinlabel{\tiny{$j$}} at 36 17
    \pinlabel{\tiny{$n$}} at 23 11
   \endlabellist
   \Mtwoa{}} & = 
    {\labellist
       \pinlabel{\tiny{$i$}} at 8 32
    \pinlabel{\tiny{$j$}} at 38 32
     \pinlabel{\tiny{$n$}} at 23 36
   \endlabellist
   \Mfoura{}} = \, \delta_{i,j}\\
     {\labellist
     \pinlabel{\tiny{$i$}} at 8 32
    \pinlabel{\tiny{$j$}} at 38 32
    \pinlabel{\tiny{$n$}} at 23 36
   \endlabellist
   \Mthreea{}} & \rightarrow  
   {\labellist
     \pinlabel{\tiny{$j$}} at 7 21
    \pinlabel{\tiny{$i$}} at 38 22
    \pinlabel{\tiny{$n$}} at 23 39
   \endlabellist
   \loopthree{}}
   = \, \omega^{-n} \delta_{i,[j+1]_{\tiny{N}} } \\
    {\labellist
        \pinlabel{\tiny{$i$}} at 6 17
    \pinlabel{\tiny{$j$}} at 38 17
    \pinlabel{\tiny{$n$}} at 23 11
   \endlabellist
   \Monea{}} & \rightarrow 
   {\labellist
     \pinlabel{\tiny{$j$}} at 5 25
    \pinlabel{\tiny{$i$}} at 36 25
    \pinlabel{\tiny{$n$}} at 21 8
   \endlabellist
   \loopone{}}
    = \, \omega^{n} \delta_{j,[i+1]_{N} } 
\end{align}

\begin{equation}
\label{rKashaev}
  \langle i,j|r(m,n)|k,l\rangle=V_{i,j-m,k-n,l}(\omega)\omega^{k-l-n+(k-i-n)m}
\end{equation}
where
\begin{align}
&  V_{i,j,k,l}(\omega):= \nonumber \\
 & \frac{N\hev{N}{\prin{N}{j-i-1}+\prin{N}{l-k}}
    \hev{N}{\prin{N}{i-l}+\prin{N}{k-j}}}{
    (\bar\omega)_{\prin{N}{j-i-1}}(\omega)_{\prin{N}{i-l}}
    (\bar\omega)_{\prin{N}{l-k}}(\omega)_{\prin{N}{k-j}}} \,
\end{align}
and $[k]_n = k\,\, {\rm mod} \,\,n$, $\theta_n(k) = \delta_{k,[k]_n}$, $(x)_n = \prod_{k=1}^n(1-x^k)$.

As it was said before to define the Kashaev invariant we base our calculations on $(1,1)$-tangle and it is a matrix invariant, which is equal to identity matrix $N\times N$ multiplied with Jones polynomial colored with $(m+1)$-dimensional representation $L_{m}$ (\ref{repnot}) (corresponding to the Young diagram $[m]$) evaluated at a point $q=\omega$:

\begin{equation}
    \langle K \rangle_{N,m} = J_{[m]} (\omega) I_N.
    \label{kashaev}
\end{equation}

\noindent This result is the conjecture based on calculations of different examples. Even though Kashaev constructed \R-matrix that is not based on representations of \uqsltwo{} the resulting polynomial coincides with Jones polynomial. 

\section{Conclusion}
 In this letter we considered different invariants of knots and links at roots of unity. Among them are colored Jones and HOMFLY-PT polynomials (\ref{homfly}) evaluated at roots of unity, ADO or colored Alexander invariants and their generalization and the Kashaev invariants.
 
 All these invariants can be defined and calculated with the Reshetikhin-Turaev method, however to define invariants at roots of unity one needs to modify the RT method:  consider of $(1,1)$-tangles instead of knots and links and introduce the normalization coefficient (\ref{ADOnorm}, \ref{norm}).
 
 We defined ADO or colored Alexander invariants (\ref{ADOdef}). They depend on a root-of-unity variable and correspond to nilpotent representations with parameters of \uqsltwo{} at roots of unity. They are connected with Jones (\ref{ADOJ}) and Alexander (\ref{ADOA}) polynomials and according to the recent study are equivalent to Jones polynomials \cite{Sonny}.
 
 We also discussed Kashaev invariants, that are defined with \R-matrix with integer parameters that depends on a root-of-unity variable and is not defined with representations of \uqslN{}. The resulting polynomial conjecturally coincides with colored Jones polynomial (\ref{kashaev}). 
 
 This letter also contains a brief summary of definition of invariants $\P_{m,N}^{\L}(\l_i^{(j)})$ (\ref{PH}) that are the generalization of ADO invariants. They are a new type of invariants, defined for nilpotent  representations with parameters $W_{m,N}^{\l_i}$ of \uqslN{} at roots of unity. They are connected with HOMFLY-PT (\ref{PH}) and Alexander polynomials (\ref{PA1},\ref{PA2}) \cite{mila1}. The question remains whether these invariants are independent or they are equivalent to colored HOMFLY-PT polynomials.

\section*{Acknowledgements}
This work was partly supported by RFBR grant 21-52-52004.

I am extremely grateful to my scientific advisors A.~Mironov and An.~Morozov for their guidance, patience and insight. I  would also like to thank V.~Alexeev, T.~Grigoryev, S.~Mironov, A.~Morozov, A.~Sleptsov, N.~Tselousov for fruitful discussions.

\end{document}